\documentclass[12pt,thmsa]{article}
\usepackage{sw20lart}

\input{tcilatex}

\begin{document}

\author{Claudio Garola\thanks{%
Dipartimento di Fisica dell'Universit\`{a} and Sezione INFN, 73100 Lecce,
Italy. \textit{E-mail}: garola@le.infn.it}}
\title{MGP versus Kochen-Specker condition in hidden variables theories}
\date{}
\maketitle

\begin{abstract}
\textit{Hidden variables theories} for quantum mechanics are usually assumed
to satisfy the \textit{KS condition}. The Bell-Kochen-Specker theorem then
shows that these theories are necessarily contextual. But the KS condition
can be criticized from an operational viewpoint, which suggests that a
weaker condition (MGP) should be adopted in place of it. This leads one to
introduce a class of \textit{hidden parameters theories }in which
contextuality can, in principle, be avoided, since the proofs of the
Bell-Kochen-Specker theorem break down. A simple model recently provided by
the author for an objective interpretation of quantum mechanics can be
looked at as a noncontextual hidden parameters theory, which shows that such
theories actually exist.
\end{abstract}

\section{Introduction}

Kochen and Specker (1967) have shown that, for every statistical theory, a
phase space $\Omega $ of hidden states and a probability measure $\mu $ on $%
\Omega $ can be introduced such that a state uniquely determines the values
of all observables and the statistical predictions of the theory are
reproduced. In this broad sense, therefore, \textit{hidden parameters
theories} exist for any statistical theory, hence for quantum mechanics (QM).

According to the standard interpretation, however, QM also yields
predictions for properties of individual samples of general physical systems
(briefly,\textit{\ individual systems}, or \textit{physical objects}). For
instance, one says that the values of mutually compatible observables can be
measured simultaneously on an individual system. Yet, it is well know that a
number of difficulties occur when trying to interpret the statistical
predictions in terms of individual systems, so that some scholars foster a
statistical interpretation of QM only. But if one accepts that also
individual systems enter into play in the interpretation of QM, some further
conditions have to be imposed on hidden parameters theories aiming to
reproduce all the results of QM (briefly, HPTs), besides the condition that
the measure $\mu $ must allow one to recover all quantum probabilities. At
first sight, one expects that these conditions follow directly from the
standard interpretation of QM, but a deeper analysis shows that there is a
degree of arbitrariness in choosing them: for, the interpretation of a
physical theory (hence, in particular, of QM) is never complete, and new
physical conditions may actually establish new, or partially new,
interpretative assumptions.

Bearing in mind the above remark, the condition that HPTs are usually
assumed to fulfill, which constitutes a basic premise for the arguments
proving the contextuality and nonlocality of HPTs, is considered in Section
2, and some criticisms that can be raised against it are resumed.
Furthermore, it is shown in Section 3 how the above criticisms can be
avoided by weakening the KS condition: this weakening entails broadening the
class of possible HPTs and implies the remarkable result that the theorems
mentioned above do not hold in the new class, so that \textit{noncontextual}
and \textit{local} HPTs may exist in it. Finally, an example of a theory of
this kind is provided in Section 4, summarizing the model that has been
propounded in some recent papers in order to show that an objective
interpretation of QM is possible (Garola, 2002; Garola and Pykacz, 2004).

\section{The KS condition}

As anticipated in Section 1, this section focuses on the condition that is
introduced by Kochen and Specker as a basic requirement ``for the successful
introduction of hidden variables''. This condition is restated by Mermin
(1993) in a very simple form, as follows.\medskip 

\noindent \textbf{KS condition}. \textit{If a set (A, B, C, ...) of mutually
commuting observables satisfies a relation of the form f(A, B, C, ...) = 0
then the values v(A), v(B), v(C), ... assigned to them in an individual
system must also be related by f(v(A), v(B), v(C), ...) = 0.\medskip }

The existence of some arbitrariness in postulating the KS condition is
explicitly recognized by Mermin, who writes, before stating it,

\noindent ``\textit{Here is what I hope you will agree is a plausible set of
assumptions for a straightforward hidden variables theory.}''

Following Kochen and Specker, it is usually assumed that only HPTs
satisfying the KS condition can be accepted, so that the name ``hidden
variables theories'' itself understands that this condition is fulfilled.
Then, all proofs of the Bell-Kochen-Specker (briefly, Bell-KS) theorem,
which states the impossibility of constructing noncontextual hidden
variables theories for QM, use the KS condition explicitly and \textit{%
repeatedly}. Hence, the theorem is proved only for HPTs that satisfy this
condition.

Notwithstanding the almost universal acceptance of the KS condition, the
fact that it does not follow directly from QM but constitutes an additional
interpretative assumption suggests that one should inquire more carefully
about its consistence with the rest of the interpretative apparatus of QM.
Whenever this inquiry is performed, one sees that this condition seems
physically plausible, but there are serious arguments for criticizing its
repeated use in the proofs. These arguments have been discussed in a number
of papers (see, e.g., Garola and Solombrino, 1996a; Garola, 2000), and
cannot be reported here in detail. The core of the criticism, however, can
be summarized as follows. The repeated use of the KS condition leads one to
consider physical situations in which several relations of the form f(A, B,
C, ...) = 0 are assumed to hold \textit{simultaneously}, though there are
observables in some relations that do not commute with other observables
appearing in different relations. Hence, one envisages physical situations
in which several empirical physical laws\footnote{%
It is well known that every general physical theory, as QM, contains both 
\textit{theoretical} and \textit{empirical} physical laws. Intuitively, a
law is theoretical if it includes theoretical terms or (in the case of QM)
noncommuting observables. A law of this kind cannot be checked directly:
rather, it must be regarded as a \textit{scheme of laws}, from which
empirical laws (that can be directly checked in suitable physical
situations) can be deduced.} (those expressed by the relations themselves)
are assumed to be simultaneously valid though they cannot be simultaneously
checked. This sounds inconsistent with the operational philosophy of QM.

It is still interesting to observe that also the proofs of nonlocality of QM
stand on assuming particular instances of the KS condition. This must
however be recognized by direct inspection, since this assumption is not
explicit in most cases.

\section{A weaker condition for HPTs}

If the criticism to the KS condition is accepted, one can try to replace
this condition with a weaker constraint, more respectful of the operational
philosophy of QM. To this end, one can start from the basic remark that the
hidden variables taken into account by the Bell-KS theorem (in order to
disprove their existence) are supposed to determine the values of all
observables independently of the environment (\textit{noncontextual hidden
variables}). This implies that two kinds of physical situations can be
envisaged because of the existence of a compatibility relation on the set of
all observables. To be precise, if x is a physical object that is produced
in a given state by means of a suitable preparing device, an \textit{%
accessible} \textit{physical situation} is envisaged whenever x is assumed
to be detected if a measurement is done and possessing some pairwise
compatible properties, while a \textit{nonaccessible physical situation} is
envisaged whenever x is assumed to be not detected if a measurement is done
or possessing properties that are not pairwise compatible.\footnote{%
From an operational viewpoint, an accessible physical situation is
characterized by the fact that one can single out a subset (that can be
void) of physical objects possessing the desired properties when considering
a set of physical objects in a state S; indeed, this can be done by
performing a suitable measurement on every physical object in the state S
(of course, the state of the objects after the measurement might not
coincide with S). No such subset can instead be singled out if a
nonaccessible physical situation is envisaged.} Now, note that the physical
situations considered at the end of Section 2 are examples of nonaccessible
physical situations. The arguments carried out when criticizing the KS
condition therefore suggest that this condition should be weakened by
restricting its validity to accessible physical situations. Thus, one is led
to state the following \textit{Metatheoretical Generalized Principle}%
.\medskip 

\noindent \textbf{MGP}. \textit{A physical statement expressing an empirical
physical law is true in all accessible physical situations, but it may be
false (as well as true) in nonaccessible situations.\medskip }

The above principle has been proposed in a number of previous papers. Here,
however, the definition of \textit{accessible physical situation} takes into
account the possibility that the physical object be not detected, which
guarantees consistency (Garola 2002, 2003; Garola and Pykacz 2004).
Substituting MGP to the stronger KS condition implies considering a class of
HPTs that includes the class of hidden variables theories in the standard
sense. In this broader class the proofs of the Bell-KS theorem break down
(it can be seen that the same occurs for the proofs of nonlocality, see
Garola and Solombrino 1996b), since the KS condition cannot be applied.
Thus, at least in principle, \textit{noncontextual (and local) HPTs are
possible}.

It remains to show, however, that that such kind of theories actually exist.
This existence has been proved in some of the papers mentioned above by
providing a model for an interpretation of QM that is \textit{objective}, in
the sense that any conceivable property of a physical system either is
possessed or not by a sample of the system, independently of any
measurement. This model (called \textit{SR model}, since it subtends an
epistemological attitude that was called \textit{Semantic Realism} in the
above papers) actually does not mention explicitly hidden parameters, but
some elements in it can be interpreted as such. Since objectivity implies
noncontextuality, these hidden parameters are noncontextual. Moreover, one
can show that they do not satisfy the KS condition (which would be
prohibited by the Bell-KS theorem), hence they are not hidden variables in
the standard sense, but satisfy MGP. Thus, the SR model provides a sample of
noncontextual HPT.

Before coming to a brief review of the SR model, note that the fact that it
satisfies MGP instead of the KS condition illustrates the price to pay in
order to avoid the contextuality of QM: one must admit that empirical
physical laws may fail to be true whenever one considers physical situations
that are classified as nonaccessible because of QM itself. This restriction
is theoretically relevant but has no \textit{direct} empirical consequence
(it may have some \textit{indirect} empirical consequences, as predicting
that the Bell inequalities can be violated also in an objective
interpretation of QM, see, e.g., Garola and Pykacz, 2004), and constitutes a
(cheap) charge for avoiding old problems and paradoxes in the interpretation
of QM. For instance, the \textit{objectification} problem in quantum
measurement theory, which remains unsolved also in some sophisticated
generalizations of standard QM, as unsharp quantum mechanics (see, e.g.,
Busch \textit{et al.}, 1991), obviously disappears in an objective
interpretation of QM. Analogously, the Schr\"{o}dinger's cat paradox, the
Wigner's friend paradox, etc., also disappear.

\section{The SR model}

As anticipated in Section 1, this section is devoted to resume the essential
features of the SR model and to illustrate qualitatively how it may happen
that some widely accepted results, as the contextuality of QM, may fail to
hold in the interpretation of QM provided by the model. This result can be
better achieved proceeding by steps, as follows.

(i) States are neatly distinguished from physical properties in the SR
model, since they are defined, as in Ludwig (1983), by means of preparation
procedures. To be precise, a state is defined as a class of physically
equivalent preparation procedures. A physical object in a given state S is
then defined by a preparation act, performed by means of a preparation
procedure that belongs to the class denoted by S. Furthermore, pure states
are represented by vectors of a Hilbert space $\mathcal{H}$ associated with
the physical system, as in standard QM.

(ii) Properties are defined as pairs ($\mathcal{A}_{0}$,$\Delta $), where $%
\mathcal{A}_{0}$ is a \textit{measurable physical quantity} (briefly, 
\textit{observable}) and $\Delta $ a Borel set on the real line, as in
standard QM. But each observable $\mathcal{A}_{0}$ is obtained from an
observable $\mathcal{A}$ of standard QM by adding to the spectrum $\Sigma $
of $\mathcal{A}$ a \textit{no-registration} position a$_{0}$ associated to a
``ready'' state of $\mathcal{A}_{0}$. The result a$_{0}$ is then accepted as
a possible outcome in a measurement of $\mathcal{A}_{0}$, so that also ($%
\mathcal{A}_{0}$,$\left\{ \text{a}_{0}\right\} $) is considered as a
possible property of the physical object x on which the measurement is
performed. Hence, obtaining a$_{0}$ is not interpreted as a failure in
detecting x because of a lack of efficiency caused by the flaws of the
concrete instrument, but as the registration of an intrinsic feature of x
(intuitively, \textit{x is such that it cannot move the ``ready'' state of }$%
\mathcal{A}_{0}$\textit{\ into a new state}).

(iii) For every Borel set $\Delta $, the property ($\mathcal{A}_{0}$,$\Delta 
$) is represented by the same (orthogonal) projection operator that
represents ($\mathcal{A}$,$\Delta \backslash \left\{ \text{a}_{0}\right\} $)
in standard QM (equivalently, ($\mathcal{A}$,$\Delta $), since a$_{0}$ does
not belong to the spectrum $\Sigma $ of $\mathcal{A}$). Therefore, whenever a%
$_{0}\in \Delta $ the properties ($\mathcal{A}_{0}$,$\Delta $) and ($%
\mathcal{A}_{0}$,$\Delta \backslash \left\{ \text{a}_{0}\right\} $) are
represented by the same projection operator (similarly, if a$_{0}\notin
\Delta $, ($\mathcal{A}_{0}$,$\Delta $) and ($\mathcal{A}_{0}$,$\Delta \cup
\left\{ \text{a}_{0}\right\} $) are represented by the same projection
operator), though they are physically different (for instance, ($\mathcal{A}%
_{0}$,$\left\{ \text{a}_{0}\text{, a}_{k}\right\} $) is the property ``being
not detected or having value a$_{k}$ of $\mathcal{A}$'', while ($\mathcal{A}%
_{0}$,$\left\{ \text{a}_{k}\right\} $) is the property ``being detected and
having value a$_{k}$ of $\mathcal{A}$''). Thus, not only physically
equivalent, but also physically inequivalent properties are represented by
the same mathematical object. In this sense we say that \textit{the
representation of properties is not bijective in the SR model}.\footnote{%
The SR model has been recently simplified, assuming that a property ($%
\mathcal{A}_{0}$,$\Delta $) has a mathematical representation (a projection
operator) only if a$_{0}\notin \Delta $ (Garola and Pykacz, 2004). In this
case, every projection operator corresponds to a property (in absence of
superselection rules), but not all properties have a mathematical
counterpart. However, the conclusions at the end of this section hold true
also in the new version of the model.}

(iv) A binary relation of \textit{commeasurability} is defined on the set of
properties, as follows: two properties F$_{1}$ and F$_{2}$ are commeasurable
iff an observable $\mathcal{A}_{0}$ exists, the measurement of which
provides a simultaneous measurement of F$_{1}$ and F$_{2}$. Furthermore,
commeasurable properties are assumed to be represented by commuting
projection operators, as in standard QM.

(v) One introduces \textit{accessible} and \textit{nonaccessible} physical
situations according to the scheme introduced in Section 3. To be precise,
one says that an accessible physical situation is considered whenever a
given physical object x in a state S is assumed to be detected and to
possess some pairwise commeasurable physical properties; one says that a
nonaccessible physical situation is considered whenever x is assumed to be
not detected or to possess properties that are not pairwise commeasurable.
Hence, in particular, a nonaccessible physical situation is considered
whenever the outcome a$_{0}$ is assumed to occur. Moreover, \textit{%
properties corresponding to the same projection operator are not physically
distinguishable in an accessible physical situation.}

(vi) The probability that a given physical object in a given state possesses
a given property can be evaluated by referring to the representations of
states and properties and using the rules of standard QM \textit{in all
accessible physical situations}. Hence the mathematical apparatus and the
statistical predictions of QM are preserved in such situations.

(vii) For every physical object x, all properties are objective in the sense
specified in Section 3, that is, they are possessed or not possessed by x
independently of any measurement. Thus, for every physical situation and
property F = ($\mathcal{A}_{0}$,$\Delta $), one can associate a value v(F) =
1 (alternatively, v(F) = 0) to F if F is possessed (alternatively, not
possessed) by x. Because of objectivity, properties can then be considered
as hidden parameters, taking values 0 or 1 (a \textit{hidden pure state} can
then be defined as an assigment of values to all properties of the system).
These parameters are necessarily noncontextual, since contextuality would
imply nonobjectivity.

(viii) Let P, Q, R, ..., be commuting projection operators, and let us
consider an empirical physical law of standard QM expressed by a relation of
the form f(P, Q, R, ...) = 0 (which is a special case of the relation f(A,\
B,\ C,\ ...) = 0 considered in Section 2, where A, B, C, ... are Hermitian
operators). According to the SR model, P, Q, R, ... do not correspond
bijectively to physical properties. Hence, if F, G, H, ... are properties
represented by P, Q, R, ... respectively, one cannot generally assert that
the values of F, G, H, ... are related by f(v(F), v(G), v(H), ...) = 0 if F,
G, H, ... are not suitably chosen. But if one considers an accessible
physical situation, properties represented by the same projection operator
are physically indistinguishable, the choice of F, G, H, ... is irrelevant,
and one expects that the values of F, G, H, ... satisfy the above relation.
Thus, the quantum law f(P,\ Q,\ R,\ ...) = 0 is fulfilled in all situations
in which one can actually test it, it may be violated in those situations
that are not accessible to experience. It follows that the hidden parameters
(properties) do not satisfy the KS condition in the SR model, but they
satisfy MGP, as anticipated in Section 3.

\section{Some remarks on locality}

It has been already noted at the end of Section 2 that special cases of the
KS-condition are understood in all existing proofs of nonlocality of QM.
Thus, one may wonder whether the substitution of this condition with MGP
also allows one to avoid nonlocality. It has been proven in a number of
papers that the answer is positive (see, e.g., Garola and Solombrino, 1996b;
Garola and Pykacz, 2004).

Furthermore, the SR model constitutes an example of HPT in which locality
holds as a consequence of objectivity. It is then interesting to compare it
with some different attempts to introduce local hidden variables of the kind
envisaged by Bell in his original paper on the EPR paradox (Bell, 1964), but
avoiding the contradiction with quantum predictions pointed out by the
original Bell inequality and by all Bell-type inequalities derived later.
This comparison has been briefly carried out in the paper in which the SR
model was propounded, and leads one to the conclusion that the aforesaid
attempts are basically different from the SR model, though there are
similarities that can mislead the reader. More precisely, all local hidden
variables theories implicitly require that the hidden variables satisfy
constraints that are equivalent to special cases of the KS condition, so
that they imply the Bell inequality, hence contradict QM. In order to avoid
this contradiction a \textit{quantum detection efficiency} can be introduced
which makes it impossible to discriminate between QM and local hidden
variables theories on the basis of the existing experimental results (see,
e.g., Clauser and Horne, 1974; Fine, 1989; Szabo, 2000; Garuccio, 2000).
But, of course, further experiments with higher efficiencies could
invalidate this kind of theories if the results predicted by QM were
obtained. On the contrary, the SR model introduces local hidden parameters
that do not satisfy the KS condition, so that they do not imply any
contradiction with QM within accessible physical situations: thus, it cannot
be disproved by empirical tests. The role of the SR model is indeed purely
theoretical: it aims to show that an objective (hence noncontextual and
local) and physically reasonable interpretation of QM is possible,
contradicting deeply-rooted beliefs and helping to avoid a number of
paradoxes. Besides this, it also suggests how QM can be embodied, at least
in principle, into a more general objective theory (Garola, 2003).

\vspace{1in}

\noindent \textbf{REFERENCES}

\bigskip

{\small \noindent Bell, J. S. (1964). On the Einstein Podolski Rosen
paradox, \textit{Physics,} \textbf{1}, 195-200.}

{\small \noindent Busch, P., Lahti, P. J., and Mittelstaedt, P. (1991). 
\textit{The Quantum Theory of Measurement}, Springer-Verlag, Berlin.}

{\small \noindent Clauser, J. F., and Horne, M. A. (1974). Experimental
consequences of objective local theories, \textit{Physical Review D}, 
\textbf{10}, 526-535.}

{\small \noindent Fine, A. (1989). Correlations and efficiency; testing the
Bell inequalities, \textit{Foundations of Physics}, \textbf{19}, 453-478.}

{\small \noindent Garuccio, A. (2000). The evolution of the concept of
correlation function in the researches on Einstein locality, in \textit{The
Foundations of Quantum Mechanics. Historical Analysis and Open
Questions-Lecce 1998}, C. Garola and A. Rossi, eds., Scientific Worlds,
Singapore, pp 219-231.}

{\small \noindent Garola, C. (1999). Semantic Realism: A new philosophy for
quantum physics, \textit{International Journal of Theoretical Physics,} 
\textbf{38}, 3241-3252.}

{\small \noindent Garola, C. (2000). Objectivity versus nonobjectivity in
quantum physics, \textit{Foundations of Physics,} \textbf{30}, 1539-1565.}

{\small \noindent Garola, C. (2002). A simple model for an objective
interpretation of quantum mechanics, \textit{Foundations of Physics}, 
\textbf{32}, 1597-1615.}

{\small \noindent Garola, C. (2003). Embedding quantum mechanics into an
objective framework, \textit{Foundations of Physics Letters}, \textbf{16},
605-612.}

{\small \noindent Garola, C., and Pykacz, J. (2004). Locality and
measurements within the SR model for an objective interpretation of quantum
mechanics, \textit{Foundations of Physics}, \textbf{34}, 449-475.}

{\small \noindent Garola, C., and Solombrino, L. (1996a). The theoretical
apparatus of semantic realism: A new language for classical and quantum
physics, \textit{Foundations of Physics,} \textbf{26}, 1121-1164.}

{\small \noindent Garola, C., and Solombrino, L. (1996b). Semantic realism
versus EPR-like paradoxes: The Furry, Bohm-Aharonov an Bell paradoxes, 
\textit{Foundations of Physics,} \textbf{26}, 1329-1356.}

{\small \noindent Ludwig, G. (1983). \textit{Foundations of Quantum
Mechanics I}, Springer-Verlag, Berlin.}

{\small \noindent Kochen, S., and Specker, E. P. (1967). The problem of
hidden variables in quantum mechanics, \textit{Journal of Mathematical
Mechanics}, \textbf{17}, 59-87.}

{\small \noindent Mermin, N. D. (1993). Hidden variables and the two
theorems of John Bell, \textit{Reviews of Modern Physics, }\textbf{65},
803-815.}

{\small \noindent Szabo, L. E. (2000). On Fine resolution of the EPR-Bell
problem, \textit{Foundations of Physics}, \textbf{30}, 1891-1909.}

\end{document}